\documentclass[12pt]{article}
\usepackage{graphicx}
\usepackage{epsf,amsmath,bbold,amsfonts,stmaryrd}
\usepackage{mathrsfs}
\usepackage{appendix}
\usepackage{amssymb}
\usepackage{float}
\usepackage{color} 
\usepackage[colorlinks]{hyperref}
\hypersetup{pageanchor=false,citecolor=red,urlcolor=red}
\usepackage{indentfirst}
\usepackage[numbers,square,comma,sort&compress,merge]{natbib}
\usepackage{multibib}

\def\a{\alpha}

\def\m{\mu}
\def\n{\nu}

\def\gg{$SU(3)\times SU(2)\times U(1)$}

\def\be{\begin{equation}}
\def\ee{\end{equation}}

\def\bea{\begin{eqnarray}}
\def\eea{\end{eqnarray}}

\def\ba{\begin{array}}
\def\ea{\end{array}}

\def\bc{\begin{center}}
\def\ec{\end{center}}

\def\bl{\begin{flushleft}}
\def\el{\end{flushleft}}

\def\br{\begin{flushright}}
\def\er{\end{flushright}}

\def\bi{\begin{itemize}}
\def\ei{\end{itemize}}

\def\bt{\begin{tabular}}
\def\et{\end{tabular}}

\newtheorem{question}{Question}
\def\bq{\begin{question}}
\def\eq{\end{question}}

\newtheorem{definition}{Def}
\def\bd{\begin{definition}}
\def\ed{\end{definition}}

\newtheorem{answer}{Answer}
\def\ban{\begin{answer}}
\def\ean{\end{answer}}

\newtheorem{possibleanswer}{Possible answer}
\def\bpa{\begin{possibleanswer}\normalfont}
\def\epa{\end{possibleanswer}}

\newtheorem{theorem}{Theorem}
\def\bth{\begin{theorem}}
\def\eth{\end{theorem}}

\begin{document}

\begin{titlepage}
\vspace{5cm}

\vspace{2cm}

\begin{center}
\bf \Large{Gauge coupling unification without leptoquarks}

\end{center}

\begin{center}
{\textsc {Georgios K. Karananas, Mikhail Shaposhnikov}}
\end{center}

\begin{center}
{\it Institute of Physics\\
Laboratory of Particle Physics and Cosmology\\
\'Ecole Polytechnique F\'ed\'erale de Lausanne (EPFL)\\ 
CH-1015, Lausanne, Switzerland}
\end{center}

\begin{center}
\texttt{\small georgios.karananas@epfl.ch} \\
\texttt{\small mikhail.shaposhnikov@epfl.ch} 
\end{center}

\vspace{2cm}

\begin{abstract}
We propose an  interpretation of the gauge coupling unification scale which is not  related to any new particle threshold. We revisit Grand Unified Theories and show that it is possible to completely eliminate the scalar as well as vector leptoquarks from the particle physics spectrum. As a consequence, in our approach the gauge hierarchy problem is put on different grounds, and the proton may be absolutely stable. In order to achieve that, we employ a number of nonlinear gauge-invariant constraints which only affect the superheavy degrees of freedom. We illustrate our considerations in a model based on the SU(5) group, with the generalization to other groups being straightforward. We discuss how scale or conformal invariance may be added to our proposal.

\end{abstract}

\end{titlepage}

\section{Introduction and motivation}

Arguably, the biggest triumph of utilizing the gauge principle is the successful description of the strong, weak and electromagnetic interactions in the context of a self-consistent theory based on the groups \gg, the Standard Model (SM) of particle physics. This framework has been remarkably successful, for it made it possible to explain a vast number of phenomena at the subatomic level, and moreover, all the particles that it predicts have now been discovered.  

Nevertheless, the indications that the SM is not the final theory of Nature are compelling. To start with, it fails to explain the neutrino masses and oscillations, it lacks a candidate for dark matter and does not incorporate a mechanism for the baryon asymmetry of the Universe. In addition, the SM is plagued by issues of purely theoretical nature. These are the strong CP, as well as the hierarchy and cosmological constant problems. The last two are related to the failure of (naive) dimensional analysis, such that unusually big fine-tunings are required in order to reconcile the predictions of the theory with the experimental data. Moreover, the presence of Landau pole related to the $U(1)$ symmetry and the Higgs sector, puts the consistency of the theory in jeopardy. Bear in mind though that this problem manifests itself at very high energies, above the Planck scale $M_\text{Pl}$, and might be resolved in the context of quantum gravity. 

Apart from the aforementioned shortcomings, when the SM is considered from the model-building point of view, it is in a sense unattractive due to its arbitrariness. Let us be more specific.  First, the way matter fields arrange themselves into different group representations appears to be random. Moreover, there are many free parameters that are  unrelated to each other. Finally, the experimental fact that the electric charge is quantized requires an explanation, since the $U(1)$ group is Abelian.\footnote{It should be noted though that a possible solution may be related to the cancellation of anomalies, see for example~\cite{Geng:1988pr,*Geng:1989un,Babu:1989tq,Foot:1990uf}.}

Many years ago it was realized that some of the SM problems could potentially be resolved once we require that it be the low energy limit of a gauge theory that enjoys invariance under a larger group $G$. This naturally led to \emph{Grand Unification}, i.e. to the hypothesis that, above a certain energy threshold, the strong and electroweak interactions are actually one and the same force. 

The models in which this is achieved---the \emph{Grand Unified Theories} (hereafter GUTs)---have attracted considerable attention and have been extensively studied throughout the years. For more details on various aspects of these theories, the interested reader is referred to the classic review by Langacker~\cite{Langacker:1980js}. Let us note that the most important GUTs are the Pati-Salam model based on $SU(4)\times SU(2)\times SU(2)$~\cite{Pati:1973uk}, the Georgi-Glashow $SU(5)$ theory~\cite{Georgi:1974sy}, as well as the $SO(10)$ unification proposed by Georgi~\cite{Georgi_SO(10)} and by Fritzsch and Minkowski~\cite{Fritzsch:1974nn}. 
 
A spectacular aspect of Grand Unification is that matter fields (quarks and leptons) can be placed neatly into multiplets of the gauge group $G$, something that yields nontrivial relations between their masses~\cite{Buras:1977yy,Georgi:1979df}. Yet another interesting point about GUTs is that in their context it is possible to predict the weak mixing angle $\sin^2\theta_W$~\cite{Georgi:1974yf}, or to put in other words, achieve the famous gauge-coupling unification. Therefore, due to the presence of a larger symmetry, the various parameters that in the SM appear to be free or arbitrary, might now be subject to constraints relating them. 

Note, however, that the price to pay for having unification is quite high. The introduction of the extra bosons needed to gauge the larger symmetry---called leptoquarks and denoted collectively by $X$ and $Y$---has a number of consequences. Since they are charged under color as well as flavor, they couple both to leptons and quarks. An aftermath of this fact is that the baryon (and lepton) numbers are not conserved like in the SM. As a result, one of the main predictions of these theories is baryon- (and lepton-) number violating processes, the most significant being the leptoquark-mediated proton decay. 
To satisfy the experimental constraints concerning the lifetime of the proton, it is necessary that these new vector bosons be superheavy, with masses $M_{X,Y}\gtrsim 10^{16}$~GeV~\cite{Georgi:1974yf,Buras:1977yy,Langacker:1980js}, i.e.  many orders of magnitude heavier than the SM particle content.\footnote{Actually, for the minimal $SU(5)$ GUT, it is the longevity of the proton that practically ruled out the theory.} 

This requirement is the core of the infamous gauge hierarchy problem, first pointed out by Gildener~\cite{Gildener:1976ai}. The Higgs mass is extremely sensitive to radiative corrections, so if such fields are present, we would expect that its mass be of the order of $M_{X,Y}$. It is then clear that in order to reconcile the theoretical value with the observed one, the relevant parameters have to be adjusted enormously, and in many orders of perturbation theory.

At this point we should mention that there have been numerous attempts to address the hierarchy problem, many of which require the existence of new dynamics above the electroweak scale. Among the most interesting ones are low-energy supersymmetry~\cite{Fayet:1974pd,*Fayet:1977yc,*Witten:1981nf,*Dimopoulos:1981zb,*Ibanez:1981yh}, composite Higgs models~\cite{Weinberg:1975gm,*Weinberg:1979bn,Susskind:1978ms} and large extra dimensions~\cite{ArkaniHamed:1998rs,Randall:1999ee}. However, so far there is no sign of New Physics up to the energies which are accessible to particle physics experiments.

On the other hand, there is a plethora of arguments why an intermediate \emph{particle physics scale} at energies between the electroweak and Planck scales might not be really needed, see~\cite{Shaposhnikov:2007nj}.  In fact, the observational puzzles of the SM can potentially be resolved with the presence of new physics only at the aforementioned energies. Since $M_\text{Pl}$, being related to gravitational interactions which are mediated by the massless graviton, is qualitatively different from the particle physics scales, it may not necessarily be associated with supermassive degrees of freedom. This way, the Higgs mass may be stable against radiative corrections due to the absence of diagrams with heavy particles running in the loops~\cite{Bardeen:1995kv,Vissani:1997ys,Shaposhnikov:2007nj,Farina:2013mla}.

Our purpose in this Letter is to argue that there might be an alternative way for a GUT symmetry to be realized, which is different from spontaneous symmetry breaking. Even though the latter mechanism plays the central role in the SM, it might not necessarily be the case that Nature repeats itself at the gauge coupling unification scale. We will illustrate that by employing a number of appropriate nonlinear \emph{gauge-invariant constraints}, it is indeed possible to achieve unification and simultaneously eliminate completely the heavy particles. 

Let us note that in the conventional treatment of GUTs, there exists a built-in mechanism for suppressing, but not eliminating, the supermassive states: as long as they are sufficiently massive and we are working at energies well below their masses (equivalently the unification scale), they can be integrated out by virtue of their equations of motion. From the low-energy perspective, the theory looks the same as the ``constrained'' one constructed in this work. However the implications for the hierarchy problem can be drastically different. The former is an effective field theory whose cutoff is a physical scale where the particles that UV-complete the theory exist; thus, the power-like divergencies cannot be discarded and will eventually give sizeable contributions to the Higgs mass. On the contrary, the cutoff scale of the latter theory is not related to the presence of any new states, so the Higgs mass does not receive important corrections and can be naturally small.

Meanwhile,  all the fields associated with the SM are not affected and in addition,  all the elegant characteristics of GUTs, such as the group structure of the fermionic multiplets, the prediction of $\sin^2\theta_W$, and the charge quantization, will be retained. For simplicity, we develop our idea by considering the most economic GUT based on $SU(5)$, since the generalization to other groups is quite straightforward.

We also speculate about the potential implications of global scale (or conformal) invariance in this setup. It is quite attractive to assume that the scales we observe in Nature are all related to each other and are generated via the spontaneous breaking of these symmetries~\cite{Bardeen:1995kv,Meissner:2006zh,Shaposhnikov:2008xb,Shaposhnikov:2008xi}. In practice, this is achieved by letting the different scales be sourced by the vacuum expectation value of a scalar degree of freedom, the dilaton. This approach allows us to keep the quantum corrections to the Higgs mass under control and without fine-tuning, as long as two requirements are met. First, there should not be any contribution from superheavy particles, something that resonates nicely with our approach here. Second, the regularization prescription must preserve the symmetry of the system~\cite{Englert:1976ep,Shaposhnikov:2008xi,Gretsch:2013ooa}. However, for this to be possible, the requirement of renormalizability has to be abandoned.\footnote{\label{foot:renorm}This should not be considered to be a major drawback, since realistic theories contain gravity which in any case is nonrenormalizable.}

The outline of this article is as follows. In Sec.~\ref{guts_old}, we briefly review some basics about GUTs. In Sec.~\ref{sec:elim_lept}, we discuss what are the appropriate constraints in order for the heavy particles to be absent from the spectrum of the theory. In Sec.~\ref{sec:SI}, we conjecture about the potential role of scale or conformal symmetry in our setup. Our conclusions are presented in Sec.~\ref{sec:concl}. 

\section{GUTs: a reminder}
\label{guts_old}

The main idea behind the canonical $SU(5)$ model~\cite{Georgi:1974sy}, is as follows. First, the 15 SM fermions of each generation are placed in the $\mathbf{5^*}$ and $\mathbf{10}$ representations of the group. Then, two scalar fields are introduced, one belonging to the adjoint ($\mathbf{24}$) and the other to the fundamental ($\mathbf{5}$). In what follows we denote them by $\Sigma$ and $H$, respectively. Their role is to effectuate the chain of spontaneous symmetry breaking 
\be
SU(5)\xrightarrow[\mathbf{24}]{}SU(3)\times SU(2)\times U(1) \xrightarrow[\mathbf{5}]{} SU(3)\times U(1) \ .
\ee
The scalar field in the adjoint representation of the group is expressed as ($\text{a}=1,\ldots,24$)
\be
\label{sigma_field_adj}
\Sigma= \Sigma_\text{a} T_\text{a} \ ,
\ee
with $T_\text{a}$ the generators, and summation over repeated indices is understood. For the first part of the breaking pattern to take place, this field should acquire a vacuum expectation value proportional to the hypercharge generator $T_{24}$, i.e.
\be
\label{vev_hyp}
\langle\Sigma\rangle=v_{GUT}\,\text{diag}(1,1,1,-3/2,-3/2) \ ,
\ee
where $v_{GUT}$ is a parameter with mass dimension. 

Next comes the breaking of the SM group down to $SU(3)\times U(1)$. This time, it has to be a scalar $H$---the Higgs field---in the fundamental representation of the group that gets a vacuum expectation value 
\be
\label{const_comp_2}
\langle H\rangle= \frac{v_{EW}}{\sqrt{2}}(0,0,0,0,1)^T\ ,
\ee
and gives masses to the SM particles. 

The above discussion indicates that the vacuum expectation values of the scalar fields $\Sigma$ and $H$ have to differ by approximately 13 to 14 orders of magnitude for the resulting theory to stand a chance of being phenomenologically viable. To make this point more clear, let us for concreteness consider a theory that possesses the discrete $\mathbb{Z}_2$ symmetry  $\Sigma\to -\Sigma$. The most general potential with terms up to quartic order contains seven invariants and reads
\be
\begin{aligned}
\label{potential}
V=&-\frac{1}{2}m_\Sigma^2\text{Tr}(\Sigma^2)-\frac{1}{2}m_H^2H^\dagger H+\frac{1}{4}\lambda_{\Sigma\Sigma}\left(\text{Tr}(\Sigma^2)\right)^2+\frac{15}{14}\lambda'_{\Sigma\Sigma}\text{Tr}(\Sigma^4)\\
&+\frac{1}{4}\lambda_{HH}\left(H^\dagger H\right)^2+\frac{1}{2}\lambda_{\Sigma H}\text{Tr}(\Sigma^2)H^\dagger H+ \frac{5}{3}\lambda'_{\Sigma H}H^\dagger \Sigma^2 H\ .
\end{aligned}
\ee
Here, $m_\Sigma, m_H,\lambda_{\Sigma\Sigma},\lambda'_{\Sigma\Sigma},\lambda_{HH},\lambda_{\Sigma H},\lambda'_{\Sigma H}$ are constants, and the normalization of coefficients was chosen for later convenience. Plugging~\eqref{vev_hyp} and~\eqref{const_comp_2} into the above, we find that the minimum of the potential corresponds to
\be
\begin{aligned}
&v_{GUT}^2=\frac{2(\lambda_{HH}m_\Sigma^2-(\lambda_{\Sigma H}+\lambda'_{\Sigma H})m_H^2)}{15(\lambda_{HH}(\lambda_{\Sigma\Sigma}+\lambda'_{\Sigma\Sigma})-(\lambda_{\Sigma H}+\lambda'_{\Sigma H})^2)} \ ,\\
&v_{EW}^2=\frac{2((\lambda_{\Sigma\Sigma}+\lambda'_{\Sigma\Sigma})m_H^2-(\lambda_{\Sigma H}+\lambda'_{\Sigma H})m_\Sigma^2)}{\lambda_{HH}(\lambda_{\Sigma\Sigma}+\lambda'_{\Sigma\Sigma})-(\lambda_{\Sigma H}+\lambda'_{\Sigma H})^2} \ .
\end{aligned}
\ee
The correct hierarchy between the vacuum expectation values of the fields, i.e.
\be
\label{hier_1}
\frac{v_{EW}}{v_{GUT}}\sim \mathcal O(10^{-13}-10^{-14})\ ,
\ee 
requires that
 \be
(\lambda_{\Sigma\Sigma}+\lambda'_{\Sigma\Sigma})m_H^2-(\lambda_{\Sigma H}+\lambda'_{\Sigma H})m_\Sigma^2\approx 0 \ ,
\ee
a relation that has to hold with an accuracy of 26 orders of magnitude. The lack of selection rules dictating that this should indeed be the case, is actually the very origin of the hierarchy problem. Note that once quantum corrections are taken into account, the problem becomes much worse, because a fine-tuning is needed at every order in perturbation theory~\cite{Gildener:1976ai}  (for a more recent discussion see  \cite{Vissani:1997ys,Shaposhnikov:2007nj,Farina:2013mla}). 

Unfortunately, this is not the end of the story. Out of the five components contained in $H$, there is a triplet carrying color quantum number and can mediate proton decay as well. Therefore, its mass should be by many orders of magnitude larger than the Higgs mass, which is related to the remaining doublet in $H$. This huge difference in the masses of the fields belonging to the same multiplet is highly unnatural and rather nontrivial---if at all possible---to be achieved and is known as the doublet-triplet splitting problem. There have been numerous proposals trying to resolve it~\cite{Witten:1981kv,*Masiero:1982fe,*DimopoulosErice,*Grinstein:1982um,*Srednicki:1982aj,*Inoue:1985cw}, mainly in the context of supersymmetric GUTs. 

It is well known that the minimal SU(5) is not a viable theory. The unification of the gauge couplings takes place at energies around $10^{14}$~GeV, leading to a proton lifetime of $10^{28}$ years, in contradiction with the observational constraints. In addition, the weak angle in this model is found to be $\sin^2\theta_W\approx 0.20$, which is below the experimental value. However, there are various extensions of $SU(5)$ that cure these problems, see~\cite{Langacker:1980js}.

\section{A novel way for the realization of the GUT symmetry}
\label{sec:elim_lept}

Having discussed about some of the most notorious problems related to unification, let us now show how they can be avoided. Our idea is actually simple: we use a different realization of the GUT symmetry that differs from the standard symmetry breaking.\footnote{It is well known that gauge symmetries do not actually get broken, but rather become concealed in what is called---in an abuse of language---the ``broken phase'' of a theory~\cite{Fradkin:1978dv,Frohlich:1980gj,*Frohlich:1981yi,Matveev:1983wg,Vlasov:1987vt}.} It is implemented via a set of specific gauge-invariant constraints that  project to zero  all the degrees of freedom that are not present in the SM. It should be stressed that we do not integrate them out, but rather we ``nullify'' them, and this can be carried out in a gauge-invariant manner. The implementation of our program is done in a number of steps. 

To start with, we wish to recover the SM group \gg , and at the same time eliminate the heavy scalar fields in the adjoint. Thus, we require that the eigenvalues $\sigma_i \ (i=1,\ldots,5)$ of $\Sigma^2$, which are of course gauge-invariant quantities, be equal to\,\footnote{We choose to constraint $\Sigma^2$, because we are assuming that the theory is invariant under $\Sigma=-\Sigma$. }
\be
\label{constr_1}
\sigma_1=\sigma_2=\sigma_3=v_{GUT}^2 \ ,~~~\sigma_4=\sigma_5=\frac{9}{4}v_{GUT}^2 \ ,
\ee
which from the geometrical point of view, this operation confines the theory on a specific manifold in the field-space, something which is commonly the case in nonlinear $\sigma$-models~\cite{GellMann:1960np}. When this is done, a generic field can be expressed as 
\be
\Sigma^2=U\begin{pmatrix}
\sigma_1&0&0&0&0\\
0&\sigma_2&0&0&0\\
0&0&\sigma_3&0&0\\
0&0&0&\sigma_4&0\\
0&0&0&0&\sigma_5 
\end{pmatrix} U^\dagger \ ,
\ee
with $U\in G$. The above spans the twelve-dimensional space of the  would-be Goldstones. The choice of the structure~\eqref{constr_1} is based on our intension to reproduce the SM. The most general case would correspond to five different eigenvalues for $\Sigma^2$, leading to  $U(1)^4$ gauge group. We cannot provide any sensible argument for singling out~\eqref{constr_1}, except  for the phenomenological one.

Next, to get rid of the vector leptoquarks, we impose the constraints 
\be
\label{constr_2}
\text{Tr}\left(T_\text{a} [\Sigma, D_\m \Sigma]\right)=0 \ ,
\ee
where as usual  $[a,b]\equiv ab-ba$ is the commutator, and $D_\m$ the $SU(5)$ gauge covariant derivative.\footnote{Note that this condition can be rewritten in a completely gauge-invariant way as
$$\sum_\text{a}\left(\text{Tr}\left(T_\text{a} [\Sigma, D_\m \Sigma]\right)\right)^2=0 \ ,$$
which leads to~\eqref{constr_2} in the Euclidean formulation of the theory. 
 }
By virtue of the commutation relations of $SU(5)$, it is straightforward to verify that the above relation guarantees that the heavy leptoquarks will be set to zero, together with the corresponding would-be Goldstones.  To understand why this is the case, let us note that when $\Sigma$ satisfies~\eqref{constr_1}, the only nonvanishing terms are the ones associated with the leptoquarks. Consequently, although the above constraints eliminate $X$ and $Y$, they do not affect at all the twelve SM gauge fields.\footnote{It should be noted that the constraints in~\eqref{constr_2} that make the leptoquark fields vanish, are similar in spirit to the ones derived by employing the coset construction for spontaneously broken symmetries. This technique is used to construct the low-energy effective action, using as input the symmetry breaking pattern. It was introduced initially for treating internal symmetries~\cite{Coleman:1969sm,Callan:1969sn}, and since then it has been generalized to spacetime symmetries~\cite{Ivanov:1975zq,Brauner:2014aha,Goon:2014ika}.  For various examples where there has been extensive use of this framework---especially in the case of spacetime symmetries---the interested reader is referred to~\cite{Delacretaz:2014oxa,Karananas:2015eha,Karananas:2016hrm}, and references therein.}

At this point, we should also get rid of the three supermassive scalar degrees of freedom in the Higgs five-plet. This is done by requiring that $H$ be subject to the following gauge-invariant algebraic condition 
\be
\label{constr_3}
H^\dagger \Sigma^2 H-\frac{3}{10}\text{Tr}(\Sigma^2)H^\dagger H = 0 \ .
\ee
This requirement eliminates the color triplet contained in $H$ in a brute-force manner, but leaves intact the remaining two components which are identified with the SM Higgs field. Thus in our proposal there is no doublet-triplet splitting problem,  it is replaced by the question about the origin of the condition~\eqref{constr_3}. Changing the relative coefficient in it would lead to a different phase of the theory, having nothing to do with the SM.

Upon employing the constraints presented here, cf. equations~\eqref{constr_1},~\eqref{constr_2} and~\eqref{constr_3}, we make sure that the degrees of freedom that have survived are the ones associated with the SM only. The resulting Lagrangian is that of the SM and is renormalizable. For the Higgs mass to be in accordance with observations, we find from~\eqref{potential} that we must impose
\be
\label{hig_mas_1}
m_H^2-\frac{1}{2}(\lambda_{HH}v_{EW}^2+15(\lambda_{\Sigma H}+\lambda'_{\Sigma H})v_{GUT}^2)\sim\mathcal O(10^4)~\text{GeV}^4 \ .
\ee
This relation constitutes a fine-tuning that is not explained. It is, however a \emph{technically natural} condition due to the absence of superheavy particles.

\begin{figure}[!h]
\centering
\includegraphics[scale=.4]{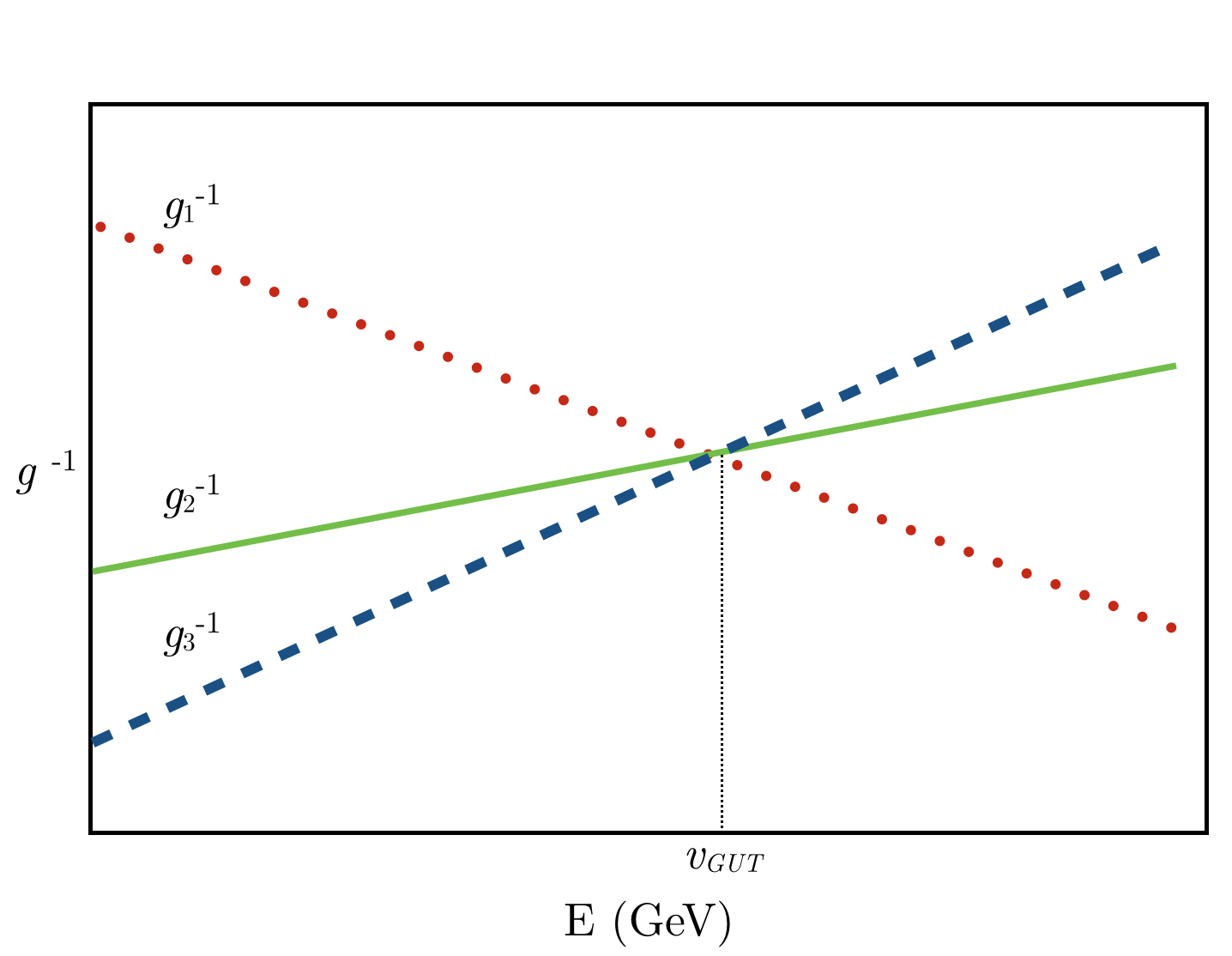}
\caption{The expected behavior (qualitatively) of the gauge couplings in our proposal. At the unification scale $v_{GUT}$, the three couplings meet but at higher energies continue to evolve independently. }
\label{fig_1}
\end{figure}

Let us comment on the energy scale $v_{GUT}$ below which the symmetry is hidden. As we have seen, $v_{GUT}$ is not related to any particle mass or new physics threshold, it serves the role of the normalisation point and can only appear through logarithms due to radiative corrections. Clearly, these effects will induce the running of the coupling constants. In our approach, the conditions that we imposed fix the relations between the SM gauge couplings at the GUT scale, see Fig.~\ref{fig_1}.  Though we only have SM degrees of freedom at all energies,  the appealing GUT features persist:  there is a nice explanation for quantum numbers of the SM fermions, the nontrivial relations between the quark and lepton masses, the prediction of the weak angle, etc. It should be noted, however, that due to the absence of the leptoquarks, the couplings do not really unify. Rather, the GUT scale is the point where they simply intersect. At the end of the next section, we will discuss how scale invariance can actually change this.

The minimal version of the theory, described above, contradicts to experiments.  Due to the absence of leptoquarks there is no problem with the decay of the proton.  However, the weak angle is the same as in the minimal $SU(5)$ (and as in any GUT containing just the SM particle content up to $v_{GUT}$), so its value is below the measured one. 

A possible way to raise the value of $\sin^2\theta_W$ and thus eliminate the tension with experiment is well known and utilizes the following idea~\cite{Hill:1983xh,Shafi:1983gz,Calmet:2009hp}. The introduction of (nonrenormalizable) higher-dimensional operators suppressed by the Planck scale can give non-negligible  contributions to the SM gauge couplings, leading to different boundary conditions for them at the GUT scale. Schematically, these operators read
\be
\mathcal O_{4+n}= \text{Tr}\left[F_{\m\n}\Sigma^k F^{\m\n} \Sigma^{n-k}\right]\ ,~~~0\le k<n \ ,~~~n>0 \ ,
\ee
where we denoted with $F_{\m\n}$ the GUT field strength, and as before $\Sigma$ is the scalar field in the adjoint of $SU(5)$. Interestingly, these operators can make $v_{GUT}$ to be pushed up and coincide with the Planck mass~\cite{Hill:1983xh,Shafi:1983gz}. This option would be quite economic and desired, indicating the unity of all known forces in Nature.\footnote{Actually, for the $SU(5)$ model under consideration here, it is enough to include only dimension five and six operators in order to achieve unification at the Planck scale~\cite{Parida:1989kg,Brahmachari:1993yn}.} Note that in our approach, the resulting theory is again renormalizable after implementing the constraints~\eqref{constr_1},~\eqref{constr_2} and~\eqref{constr_3}.

If our line of reasoning is followed, then no intermediate energy scale is present between the electroweak and Planck scales. On the one hand, this might be alarming in view of the need to go beyond the SM to explain a number of observations. On the other hand, there is no indication of New Physics at the LHC and at high precision experiments. The way out could  be the introduction of very weakly interacting particles with masses below $v_{EW}$ (ranging from keV to a few GeV). For example, three sterile Majorana neutrinos are able to simultaneously reproduce correctly the neutrino oscillation patterns, explain the baryon asymmetry of the Universe and also account for the dark matter abundance~\cite{Asaka:2005an,Asaka:2005pn,Boyarsky:2009ix}. This minimalistic scenario, the Neutrino Minimal Standard Model ($\nu$MSM), makes it possible to address a plethora of the SM shortcomings. It should be noted though that in the model based on $SU(5)$ that we study here, these right-handed neutrinos should be introduced in an ad hoc manner, like in the SM. This will not be the case if the GUT is constructed on the basis of the $SO(10)$ group, where each generation of SM matter content plus a sterile neutrino fit in the $\mathbf{16}$ representation of the group.

Before moving to the next section, it is worth taking a short detour to argue that constraints without dynamical origin are not as uncommon as it may seem at first sight. The simplest example that comes to mind is General Relativity: the covariant derivative of the metric is required to be zero in an ad hoc manner, yielding a vanishing nonmetricity tensor~\cite{Sotiriou:2008rp}. Yet another example is provided by the gravitational theory resulting from the gauging of the Poincar\'e group~\cite{Utiyama:1956sy,*Brodsky:1962,*Sciama:1962,*Kibble:1961ba}. To achieve invariance under local translations and Lorentz transformations, it is necessary to introduce the vielbein and (spin) connection, respectively. However, there is the possibility to set torsion to zero, a condition that  need not follow from the equations of motion and is practically identical to~\eqref{constr_2}. By doing so, the connection is expressed in terms of the derivative of the vielbein. It is important to note that the full Poincar\'e symmetry is still present, albeit with the use of less degrees of freedom. Alternatively, in the context of teleparallelism~\cite{DeAndrade:2000sf}, one can formulate a gravitational theory only in terms of torsion, by enforcing curvature to be zero. The list of theories which are subject to constraints does not end here. Consider the conformally coupled scalar field, which has been studied a lot throughout the years. It is well known that this specific theory is invariant under gauged Weyl rescalings, although a compensating vector field is not present. This implies that the symmetry is realized in a nontrivial manner, since the Ricci scalar is responsible for canceling the inhomogeneous piece(s) stemming from the kinetic term of the field~\cite{Iorio:1996ad}.\footnote{Actually, there exists an infinite number of higher derivative theories in which Weyl invariance is achieved due to the presence of curvature rather than a gauge field, see for example~\cite{Karananas:2015ioa}.} Obviously, this is not a coincidence, for there is a nonlinear constraint that enables one to express a certain combination of the gauge field in terms of curvature~\cite{Karananas:2015eha}.\footnote{Similar constraints exist in nonrelativistic field theories~\cite{Karananas:2016hrm}.} Once again, there is no dynamical origin for the aforementioned condition, but rather it is put by hand. It should be noted that the logic behind all these considerations is to  achieve invariance with the minimal number of compensating degrees of freedom, which is actually our strategy in this paper. 

\section{The inclusion of scale or conformal invariance}
\label{sec:SI}

In Sec.~\ref{guts_old}, we saw that for the GUT under consideration to yield a phenomenologically acceptable low energy dynamics, two scalar fields are needed. One of them, denoted by $\Sigma$ belongs to the adjoint representation of the group and is responsible for recovering the SM symmetry group. Subsequently, the Higgs field $H$ in the fundamental representation has to acquire a vacuum expectation value in order for the symmetry breaking pattern \gg$\to~SU(3)\times U(1)$, to take place. 

The most troublesome point is the lack of an underlying principle for the values of the scalars to differ by many orders of magnitude, which is even worse if we require that $v_{GUT}\sim \mathcal O(M_\text{Pl})$.  It would therefore be quite interesting to at least try and put this problem in a different context. A minimalistic way to proceed is to conjecture that all scales in the theory appear as a result of the nonlinear realization of global scale/conformal invariance. 

This can be implemented by requiring that the eigenvalues of the field  $\Sigma$ are related to a scalar field, the dilaton $\chi$, which is nothing else than the Goldstone boson of the broken scale transformations. 
Thus, eq.~\eqref{constr_1}, should be replaced by
\be
\label{constr_1_alt}
\sigma_1=\sigma_2=\sigma_3=\a\chi^2 \ ,~~~\sigma_4=\sigma_5=\frac{9\a}{4}\chi^2  \ ,
\ee
where $\a$ is a dimensionless constant. The remaining two conditions~\eqref{constr_2} and~\eqref{constr_3} that we introduced, remain the same as in previous section. It is clear that the surviving degrees of freedom in the scalar sector of the theory are the dilaton and the Higgs doublet $h$.

To construct the scale-invariant potential, we should add a quartic self-interaction for the dilaton, $\Lambda' \chi^4$, and replace the mass terms for $\Sigma$ and $H$~in eq.~\eqref{potential}, by
\be
m_\Sigma^2=\frac{15\n}{4}\a\chi^2 \ ,~~~m_H^2=\frac{15\m}{2}\a\chi^2 \ ,
\ee
with $\Lambda',\m$ and $\n$ dimensionless constants. Upon use of the aforementioned constraints, we find that the potential boils down to
\be
\label{potential_2}
V=\lambda\left(h^\dagger h -\frac{\beta}{2\lambda}\chi^2\right)^2 +(\Lambda+\Lambda')\chi^4 \ ,
\ee
where $\lambda,\beta,\Lambda$ are related to the constants appearing in~\eqref{potential}, as
\be
\begin{aligned}
&\lambda=\frac{\lambda_{HH}}{4}\ ,~~\beta=\frac{15\a}{4}(\m-\lambda_{\Sigma H}-\lambda_{\Sigma H}') \ ,\\
&\Lambda=\left(\frac{15\a}{4}\right)^2\left(\lambda_{\Sigma\Sigma}+\lambda_{\Sigma\Sigma}'-\n-\lambda_{HH}^{-1}(\m-\lambda_{\Sigma H}-\lambda_{\Sigma H}')^2\right) \ .
\end{aligned}
\ee
For scale-invariance to be broken, the potential must posses a flat direction, which corresponds to $\lambda,\beta>0$ and $\Lambda+\Lambda'=0$. Note, however, that in the presence of gravity, there exist flat directions even for $\Lambda+\Lambda'\neq0$. This in turn induces a cosmological constant term which we have to require that it be extremely small to have agreement with observations.

Note also that in order to reproduce the hierarchy between the electroweak and GUT scales, we have to impose
\be
\frac{\beta}{\a}\ll 1 \ ,
\ee
a technically natural requirement, since the dilaton has an approximate shift symmetry
in the limit $\Lambda+\Lambda' \to 0,~\beta \to 0$.  It is well known, however, that at the quantum level scale invariance is anomalous due to the introduction of a parameter with dimensions of mass during the regularization procedure. This breaks the symmetry explicitly. The resolution of this is possible by assuming that the spontaneously broken scale or conformal invariance, as well as the approximate shift symmetry mentioned above, are maintained at the quantum level. Practically, this can be implemented by the use of a subtraction scheme based on dimensional regularization with a field-dependent normalization point, that is related also to the dilaton~\cite{Englert:1976ep,Shaposhnikov:2008xi}, or field-dependent cutoff \cite{Wetterich:1987fm}. Then, the hierarchy problem is solved in a technical sense because the radiative corrections are kept under control. However, by following this procedure, the renormalizability of the theory is lost (see footnote~\ref{foot:renorm}).  

In addition to its relevance for the hierarchy problem, (global) scale invariance has a number of cosmological implications as was pointed out in~\cite{Wetterich:1987fm,Wetterich:1987fk}. Since then, there have been numerous works on this topic, see for example~\cite{Wetterich:1994bg,Shaposhnikov:2008xb,GarciaBellido:2011de,Bezrukov:2012hx,Rubio:2014wta,Karananas:2016kyt,Ferreira:2016vsc}.
In~\cite{Shaposhnikov:2008xb,GarciaBellido:2011de}, the \emph{Higgs-dilaton model}, a two-field model with a potential similar to the one in~\eqref{potential_2}, was studied in great detail. It was shown that it can account for an inflationary period in excellent agreement with the latest observational data. Interestingly, the effect of scale invariance in this setup is twofold. First, it puts constraints on the Higgs and dilaton, which during inflation are forced to move on specific trajectories (ellipses) in the field-space, something that has also been discussed in~\cite{Ferreira:2016wem}. This behavior will also be present if inflation is studied in the framework of the ``constrained GUTs'' that we introduced here. 
Second, the present day accelerated expansion of the Universe in this setup is related to dynamical dark energy whose role is played by the dilaton. This establishes a nontrivial link between the inflationary epoch and present day, in the form of testable relations between the model's observables concerning these two periods.

Let us briefly discuss the origin of the dilaton. It is in principle possible to avoid introducing this field in an ad hoc manner, since it can be associated with the determinant of the metric and thus have gravitational origin in the context of theories invariant under volume-preserving diffeomorphisms~\cite{Blas:2011ac,Karananas:2016grc}. 

\begin{figure}[!h]
\centering
\includegraphics[scale=.4]{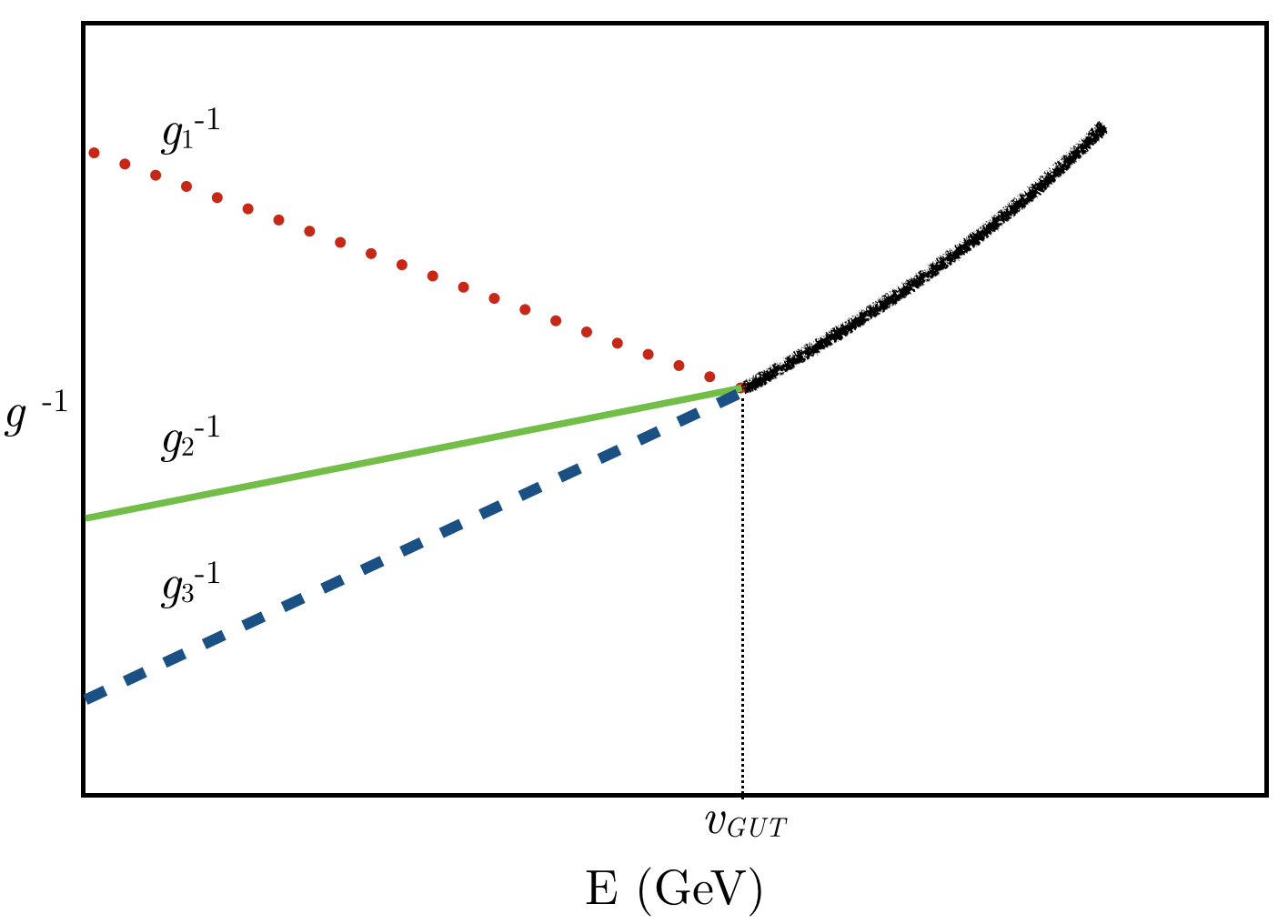}
\caption{In the presence of scale or conformal invariance, the couplings are expected to merge at $v_{GUT}$ and evolve as one at higher energies. This coincides with what happens in the traditional way of the GUT symmetry realization.}
\label{fig_2}
\end{figure}

Before concluding, we would like to speculate about the ultraviolet domain of the theory, $E\to\infty$. A possible conjecture is that this limit corresponds to a vanishing vacuum expectation value of the dilaton, $\langle\chi\rangle\to 0$, a situation that we also explored in a different context in~\cite{Karananas:2016grc}. Inspection of the constraints that we introduced in the previous section reveals that~\eqref{constr_2} and~\eqref{constr_3} are trivially satisfied, whereas~\eqref{constr_1_alt} indicates that the eigenvalues of $\Sigma$ vanish, which is equivalent to $\Sigma\to 0$. As a result, the high energy degrees of freedom are the ones related to the $SU(5)$ in the symmetric phase plus the Higgs five-plet, the SM fermions and the dilaton. In this case, it is conceivable that all three SM couplings will run together, as in the canonical GUTs, and consequently, we could potentially have asymptotically free evolution of the gauge couplings as is schematically illustrated in Fig.~\ref{fig_2}. 

\section{Concluding remarks}
\label{sec:concl}

There is no convincing reason why the dynamics governing the low and high energies should be the same, especially in view of the fact that even the motivation behind the SM in the `60s and GUTs in the `70s, is completely different. To make this point more clear, let us adopt a bottom-up perspective. At energies well below $v_{EW}$, weak interactions can be studied in the context of Fermi's theory, which is a perfectly valid description at this energy domain. However, owing to the fact that the Fermi coupling constant carries dimension GeV$^{-2}$, the theory is nonrenormalizable and its range of validity extends up to, roughly speaking, $300$~GeV. Around these energies, perturbation theory is no longer applicable and the predictions cannot be trusted anymore. An appropriate modification, or better say ultraviolet  completion, is provided by the electroweak theory, in which the massive $W^\pm$ and $Z$ bosons are the mediators of the weak interactions, and the dynamical Higgs field ensures its renormalizability. Contrary to the Fermi theory, the SM is a perfectly well-defined theory up to the energy where the Landau pole is located, which  is  well  above $M_\text{Pl}$. Thus, the approach to Grand Unification may be quite different in comparison with the SM. This is because GUTs mainly address some of the ``aesthetic'' issues of the SM and at the same time provide an economic framework in which all known forces (apart from gravity) unify at high energies.

In this article we provided a novel perspective on how gauge coupling unification may be realized. We succeeded in embedding the Standard Model of particle physics into a theory invariant under the bigger gauge group $SU(5)$, but without the presence of the leptoquarks. We showed that as long as the constraints~\eqref{constr_1},~\eqref{constr_2} and~\eqref{constr_3} are satisfied, the superheavy degrees of freedom are completely absent, so they cannot destabilize the Higgs mass. As a consequence,  the hierarchy problem is put in a different footing.
Moreover, all the successes of GUTs are passed down to the SM, providing an explanation, among others, to the quantum numbers of matter fields, the weak mixing angle, the electric charge quantization.

Clearly, there is a number of problems in our proposal, some of them similar to those of standard GUTs. First of all, the choice of symmetry is arbitrary. Although $SU(5)$ is the smaller group that contains \gg, this particular choice might not be the most economic, since for the fifteen fields in each fermionic generation of the SM, we need to employ two representations of the group. This however can be remedied by considering $SO(10)$ instead. Furthermore, as in usual GUTs, we cannot answer why there are three generations in the SM. Also, we still cannot provide an explanation regarding why the electroweak and Planck scales should differ so dramatically, though this is technically natural. While in the usual treatment of GUTs the pattern of spontaneous symmetry breaking may have dynamical origin and stem from the choice of the scalar potential, in our case the necessary constraints are postulated and look rather ad hoc. This is the weakest point of our proposal. As for the presence of scale or conformal invariance, there is a price to be paid: renormalizability has to be abandoned for the symmetry to survive at the quantum level and the resulting theory to be viable. 

\section*{Acknowledgements}

This work was supported by the ERC-AdG-2015 grant 694896.  The work of G.K.K. and M.S. was supported partially by the Swiss National Science Foundation. We thank A. Boyarsky and O. Ruchayskiy for helpful comments, and A. Monin, R. Percacci, J. Rubio and C. Wetterich for discussions.

\bibliographystyle{utphys}
\bibliography{GUTs}{}

\end{document}